\documentclass{aa}
\usepackage{graphicx}
\usepackage{natbib}
\bibpunct{(}{)}{;}{a}{}{,}

\newcommand{\bm}[1]{\mbox{\boldmath{$#1$}}}
\newcommand{\be}{\begin{equation}}
\newcommand{\ee}{\end{equation}}

\newcommand{\degrees}[1]{\ensuremath{#1^\circ}}

\begin{document}

\title{Refraction in a pulsar magnetosphere --- the effect of a variable emission height on pulse morphology}
\titlerunning{The effect of a variable emission height on  pulse morphology}
\author{P. Weltevrede\inst{1} \and B. W. Stappers\inst{2,1} \and L.J. van den 
Horn\inst{1,3} \and R. T. Edwards\inst{1}}

\date{}
\institute{Astronomical Institute ``Anton Pannekoek'', 
        University of Amsterdam,
        Kruislaan 403, 1098 SJ Amsterdam, The Netherlands 
  \and
   Stichting ASTRON, Postbus 2, 7990 AA Dwingeloo, The Netherlands   
   \and
   Institute for Theoretical Physics, University of Amsterdam, Valckenierstraat 
65, 1018 XE Amsterdam, The Netherlands   }
\offprints{P. Weltevrede, \email{wltvrede@science.uva.nl}}

\abstract{ The \cite{Petrova:Refraction} model to calculate pulse profiles
is extended to a variable emission height model to make it physically
self-consistent. In this context variable means that the emission height is
no longer considered to be the same for different magnetic field lines.  The
pulse profiles calculated using this new model seem to be less realistic due
to a focusing effect and cannot be used to fit (typical) multifrequency
pulsar observations. Apart from the focusing effect the general morphology
of pulse profiles is not greatly affected  by introducing a variable emission
height. Additional extensions of the model will be needed to be able to
fit observations, and several suggestions are made.

 }

\maketitle

\section{Introduction}

\cite{Arons:Wave} have derived the dispersion relation for three wave modes which can propagate through the plasma of a pulsar magnetosphere: the
ordinary subluminous mode (subluminous O-mode), the ordinary superluminous
mode (superluminous O-mode) and the extraordinary mode (X-mode). The X-mode
does not suffer refraction, but refraction of the subluminous O-mode can be
considerable in pulsar magnetospheres (\citealt{Barnard86}). The subluminous
O-mode cannot escape the pulsar magnetosphere due to Landau damping, so it
does not contribute directly to the observed emission.  \cite{Lyubarskii96}
has shown that the subluminous O-mode can be converted into the
superluminous O-mode -- which can escape the magnetosphere  -- by induced
scattering off plasma particles.  As pointed out by \cite{Barnard86}
refraction of the superluminous O-mode is less severe than for the
subluminous O-mode. It can, however, be important in the presence of a
transverse plasma density gradient.

For the superluminous O-mode \cite{Petrova:Refraction} (hereafter P2000) \rm shows how pulse profiles can be calculated taking into account the
transverse plasma density gradient.  This model demonstrated 
that complex profiles can be produced by a ``simple'' ring shaped emission
region (as predicted by \citealt{RS}), and thus that the wealth of observed pulse
profile shapes may be due to different magnetospheric conditions rather than
more complex emission region shapes.  Furthermore it was shown that the
observed phenomenon of high frequency core splitting could be an effect of
refraction.

The emission height is an important ingredient in calculating pulse
profiles.  The emission height is frequency dependent; i.e. there is
radius-to-frequency mapping (\citealt{Cordes78}). Plasma waves with higher
frequencies are excited closer to the star.  The observed frequency
dependence of pulse profiles is often very complex, perhaps more complex
than can be expected from just radius-to-frequency mapping. Because
refraction itself is a frequency dependent phenomenon, a more complex frequency dependence of pulse profiles can be
expected if refraction is important in pulsar
magnetospheres. Other effects that can be understood by taking into account
refraction are the occurrence of orthogonal polarization modes
(\citealt{Petrova01}) and the spectral breaks of pulsars
(\citealt{Petrova02}).

To link the observed pulse profiles to the shape of the emission region, so as
to be able to check emission theories, one must know the refractive
properties of pulsar magnetospheres.  This calls for the development of
improved refraction models.  As noted by P2000, the emission surface at one
observing frequency should be, strictly speaking, an isodensity surface of
the plasma distribution.  Yet, for simplicity, a constant emission height
(CEH) was assumed in P2000, in the expectation that the qualitative features
of profile formation would not be sensitive to that assumption. In the
present paper we do adopt a surface of constant density as required for
self-consistency of the refractive model, and we investigate the effects on
the pulse morphology. This `variable' emission height (VEH) appears to
introduce a focusing effect which causes the profiles to have
unrealistically sharp edges. As a consequence, the VEH model cannot be used
to fit multifrequency pulsar observations without relaxing additional
restrictive assumptions, a number of which are discussed at the end of the
paper.

\section{Refraction model}

\subsection{The ray equations}

\begin{figure}
\resizebox{\hsize}{!}{\includegraphics{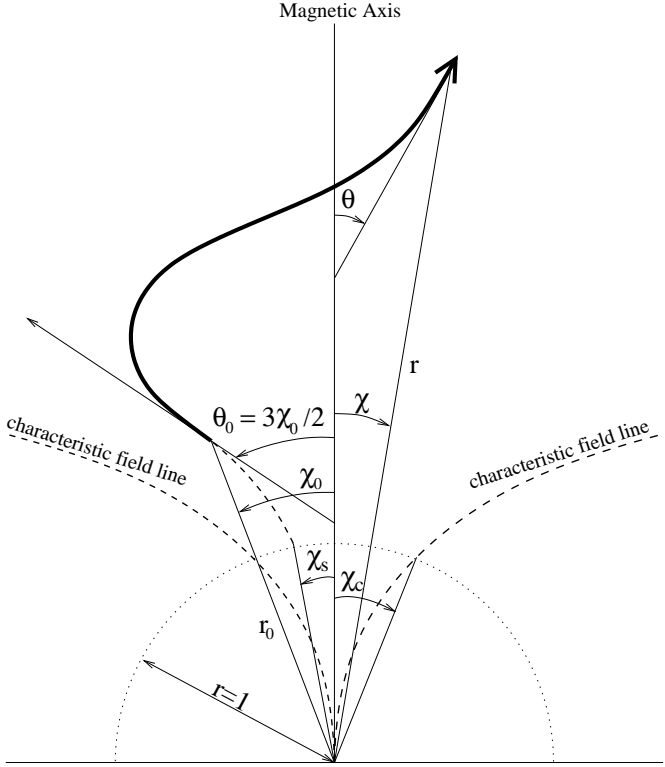}}
\caption{The ray at position $(r,\chi)$ is propagating in the direction
$\theta$. This ray was emitted at $(r_0,\chi_0)$ and the field line through
this point is indicated by $\chi_s$. The plasma density peaks at the
characteristic field lines indicated by $\chi_c$. The angles $\chi_c$ and $\chi_s$ are defined at $r=1$ and the angles $\chi_0$ and $\theta_0$ at the emission height $r_0$.} 
\label{fig:Coordinates}
\end{figure}

The refraction model below is essentially that of P2000, and we refer to
that paper for details.  The plasma distribution and the magnetic field are
assumed to be axisymmetric around the magnetic pole, so the refraction model
can be described in two dimensions.  A position on a ray trajectory is
indicated by the polar coordinates $r$ and $\chi$, and the direction along
the trajectory by $\theta$ (see Fig. \ref{fig:Coordinates}).
\label{Sect:Disp}

The geometrical optics description applies and the time evolution of these
quantities is given by the Hamilton equations.  For a highly magnetized
ultrarelativistic electron-positron plasma, which is cold in the proper
restframe, the dispersion relation has been derived by \cite{Arons:Wave} and
the associated Hamilton equations by \cite{Barnard86}.  On the condition
that the plasma flows with the same velocity for all
field lines and when rays are emitted
parallel to the local (dipolar) magnetic field, the dispersion law describing the two
ordinary wave modes can be written as (P2000)
\be
\label{DispersionLawNormalized}
\eta\left(1-\frac{4N}{f_0^2\left(1+\eta\right)^2}\right)-\frac{9}{4}\chi_0^2\gamma^2(\theta_\mathrm{n}-\chi_\mathrm{n})^2
= 0,
\ee
while from the Hamilton equations one finds
\begin{eqnarray}
\nonumber r\frac{d\chi_\mathrm{n}}{dr} &=& \frac{\chi_\mathrm{n}}{2} + 
\frac{3}{2}\frac{(1+\eta)^3\left(\theta_\mathrm{n}-\chi_\mathrm{n}\right)}{A_1}\\
\label{eq:DiffEq}r\frac{d\theta_\mathrm{n}}{dr} &=& \frac{N}{A_1f_0^2}\left[
6(1-\eta)\left(\theta_\mathrm{n}-\chi_\mathrm{n}\right) - 
A_2\frac{\partial\ln N}{\partial \chi_\mathrm{n}} 
\right],
\end{eqnarray}
where
\begin{eqnarray}
\nonumber A_1 &=&  (1+\eta)^3-\frac{4(1-\eta)N}{f_0^2}\\
\nonumber A_2 &=& \frac{4\eta(1+\eta)}{3\chi_0^2\gamma^2}.
\end{eqnarray}

The radial plasma density derivative has been omitted, because it can be neglected for the plasma density we will adopt (P2000).
The parameter $\eta$ is related to the component of the wave vector \bm{k} 
in the direction of the local magnetic field and is defined as
\be
\label{EtaDefenition} \eta = 2\gamma^2(1-n_\parallel)
\ee
with $\gamma \gg 1$ the Lorentz factor of the outflowing plasma, 
and $n_\parallel=ck_\parallel/\omega$ where $\omega$ is the frequency of the 
plasma wave. The refractive index is $n=(n_\parallel^2+n_\perp^2)^{1/2}$,
where parallel/perpendicular is with respect to the local magnetic field. 
It is assumed that $n_\parallel$ is such that $\eta\ll2\gamma^2$. 
The plasma waves are assumed to be generated close to the local Lorentz-shifted plasma 
frequency $\omega_p\sqrt{\gamma}$,
\be
\label{eq:OmegaP}
\omega_p = \sqrt{\frac{4\pi N_pe^2}{m}},
\ee
with $e$ the electron charge, $m$ the electron rest mass and $N_p$ the 
particle number density (electrons plus positrons) of the plasma. The ratio
\be
\label{fDefenition}f = \frac{\omega}{\omega_p\sqrt{\gamma}}
\ee
should then be close to unity.

\begin{figure*}
\begin{center}
  \includegraphics[width=0.75\columnwidth, angle=0]{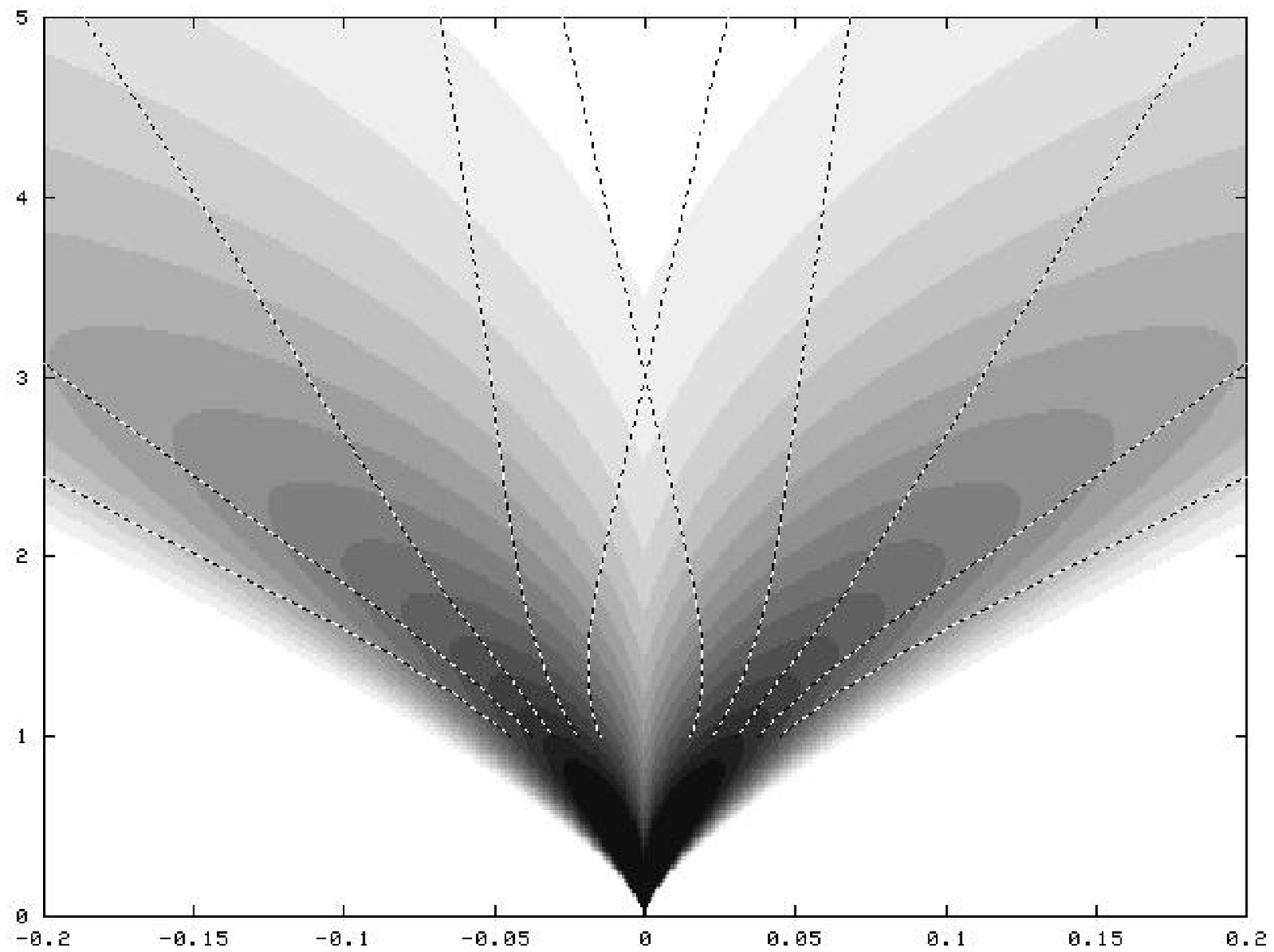}
\hspace{2cm} 
  \includegraphics[width=0.75\columnwidth, angle=0]{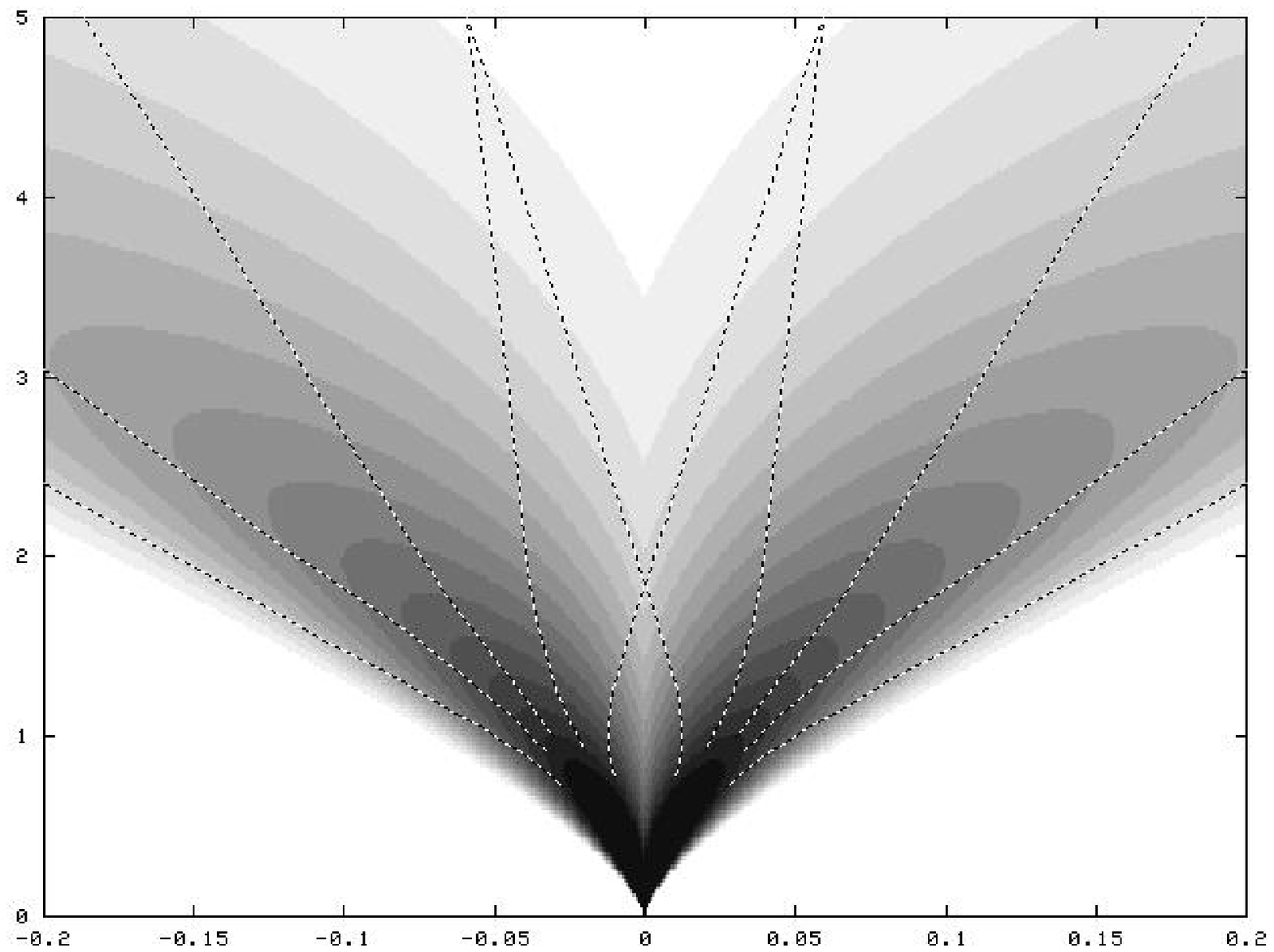}
\\
  \includegraphics[width=0.75\columnwidth, angle=0]{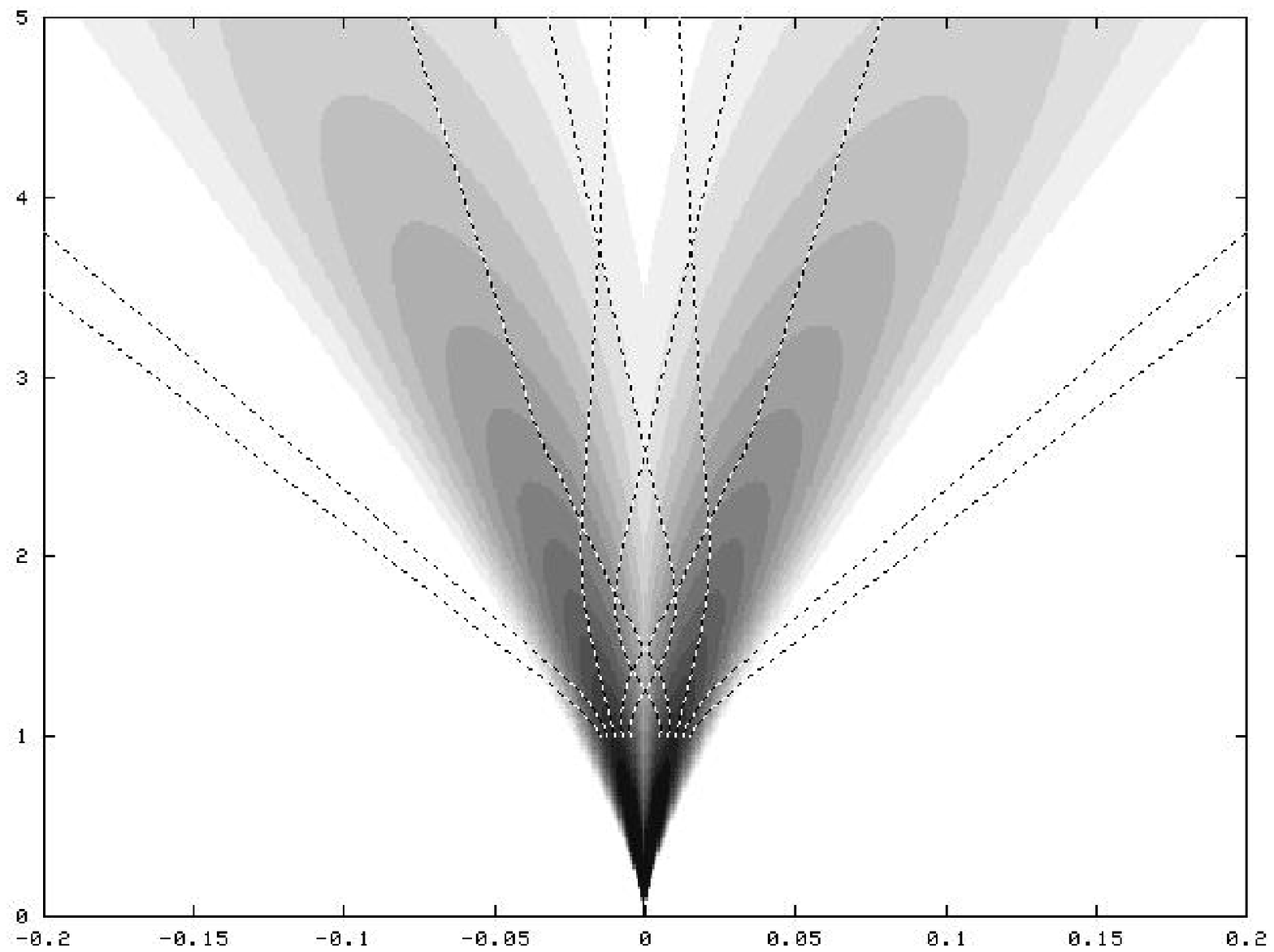}
\hspace{2cm}
  \includegraphics[width=0.75\columnwidth, angle=0]{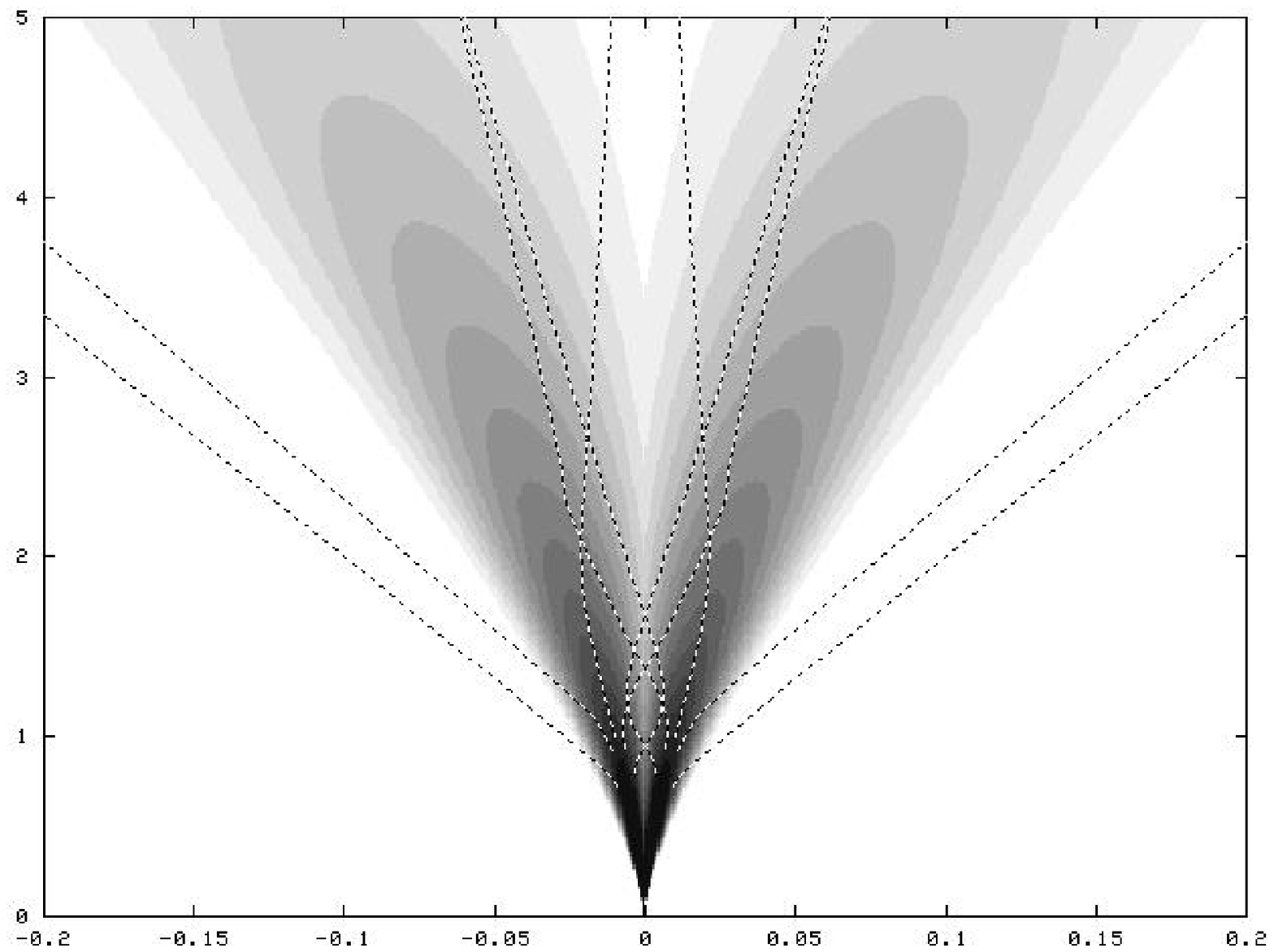}
\\
\includegraphics[width=0.3\columnwidth]{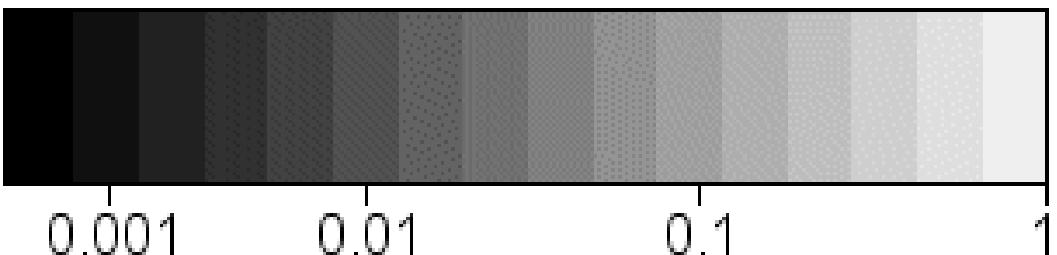}
\end{center}
\caption{\label{Fig_Refraction_Hearts}Rays (dashed) with the observing frequency are generated at the emission height and are refracted toward lower plasma densities.
The scale is in units emission height and the gray scale indicates the hollow cone plasma density. The plasma density is proportional to the numbers of the gray scale bar.
The plots on the left are for the CEH model, the plots on the right are for the VEH model, the top plots are calculated with $\chi_c=0.03$ and the bottom plots with $\chi_c=0.01$.
The other input parameters are:
$\gamma=30$, $f_0=0.5$, $\varepsilon_1 = 3$, $\varepsilon_2 = 4$.
}
\end{figure*}

Eqs. (\ref{DispersionLawNormalized}) and (\ref{eq:DiffEq}) are \em
normalized\rm, i.e. the coordinates $r$, $\chi_\mathrm{n}$ and
$\theta_\mathrm{n}$, as well as the plasma number density distribution $N$,
are normalized to their values at the emission height (so
$\chi=\chi_0\chi_\mathrm{n}$ and $\theta=\theta_0\theta_\mathrm{n}$).  The
emission height can be different for different rays as will be discussed in
Sect. 2.3, so in this context \em the \rm emission height is the
emission height of the particular ray that is being considered.  The values
of $f$ and $\chi$ at the emission height are denoted as $f_0$ and $\chi_0$
respectively.  All angles are assumed to be small compared to $1$ throughout
this paper, so the propagation direction of the plasma waves should always
be nearly parallel to the magnetic axis.

For the superluminous O-mode which is considered here, $n_\parallel$ 
cannot be larger than 1, therefore $\eta$ is required to be positive (or zero). 
At the emission height (where $\theta_\mathrm{n}=\chi_\mathrm{n}=N=1$) the solution of the 
dispersion relation follows immediately and we have for the superluminous O-mode
\be
\label{Eqn_Refr_eta0}\eta_0 = \left\{
\begin{array}{lll}
2/f_0 - 1&\mathrm{for\;\;\;}0 < &f_0 < 2\\
0&\mathrm{for\;}&f_0 \geq 2\\
\end{array}
\right.
\ee
Note that $\eta_0$ is continuous at $f_0=2$. The solution of the dispersion relation applicable
above the emission height is given by the general solution of the cubic 
(\ref{Eqn_Reft_EtaAnalytisch}).

Eqs.
(\ref{DispersionLawNormalized}) and (\ref{eq:DiffEq}) describe, to first order in $\chi_n$ and $\theta_n$, the refraction of an ordinary (both the sub- and
superluminous) plasma wave.  The two differential equations for $\chi_\mathrm{n}(r)$ and $\theta_\mathrm{n}(r)$ can be solved
numerically if $\eta(r)$
is known. As noted above, $\eta$ can be calculated analytically from the
dispersion equation.  We use a fourth order Runge-Kutta method with adaptive
stepsize control (\citealt{NR}) to solve the set of differential equations. For the plasma density distribution we adopt the hollow cone model (P2000)

\begin{figure*}[tb]
\begin{center}
\includegraphics[width=1.99\columnwidth, angle=0]{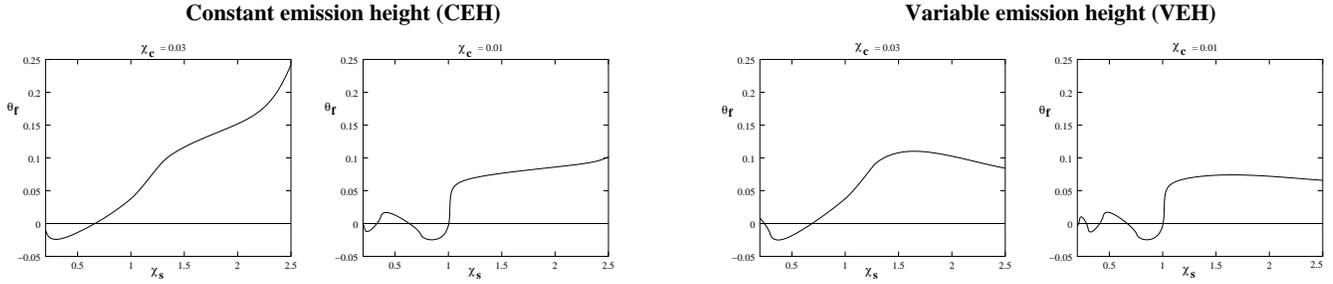}
\end{center}
\caption{\label{Fig_Refr_FRD}The final propagation direction $\theta_f$ versus $\chi_s$ for both the VEH and the CEH model,  for two values of \rm $\chi_c$. 
In these plots $\chi_s$ ranges from $\chi_c/5$ to $5\chi_c/2$ and the other parameters are the same as for Fig. \ref{Fig_Refraction_Hearts}. The final propagation direction $\theta_f$ is in radians and $\chi_s$ in units $\chi_c$.
}
\end{figure*}

\be
\label{Eqn_Refr_PlasmaDensity}
N_p = \frac{N_\star}{r^3}\exp\left(-\varepsilon\left(\frac{|\chi|-\chi_c\sqrt{r}}{\chi_c\sqrt{r}}\right)^2\right),
\ee
where $N_\star$ is the particle number density at $(r = 1,\chi=\chi_c)$.
This plasma density is Gaussian shaped around a ``characteristic field line'' indicated by $\chi_c$.
The decrease of the plasma density may be different for the inner and outer regions, so we set
\be
\label{Eqn_Refr_Epsilon}
\varepsilon =  \left\{ \begin{array}{ll}
\varepsilon_1\,\,\,\,\mathrm{for} &|\chi|\le\chi_c\sqrt{r}\\
\varepsilon_2\,\,\,\,\mathrm{for} &|\chi|\ge\chi_c\sqrt{r}\\
\end{array}
\right..
\ee
Eq. (\ref{eq:DiffEq}) do not contain $r_0$, so the whole problem is 
independent of the scaling of $r$. However, in Eq. 
(\ref{Eqn_Refr_PlasmaDensity}) $\chi_c$ is defined at $r=1$.
The normalized plasma density is given by
\be
N(\chi,r)=\frac{N_p(\chi,r)}{N_p(\chi_0,r_0)}
\ee
and its derivative with respect to $\chi_\mathrm{n}$ is
\begin{eqnarray}
\label{Eqn_Petrova_PlasmaDerivatve1}\frac{\partial\ln N}{\partial\chi_\mathrm{n}} &=& -2\varepsilon\chi_0\frac{\chi-\chi_c\sqrt{r}\,\mathrm{sign}(\chi)}{(\chi_c\sqrt{r})^2}.
\end{eqnarray}
The parameters required to calculate a single ray trajectory are
those of the plasma density distribution ($\chi_c$, $\varepsilon_1$ and $\varepsilon_2$), the plasma outflow Lorentz factor $\gamma$, the frequency of the plasma wave (expressed in $f_0$) and the start position $\chi_0$ of the ray.
Solving Eqs. (\ref{DispersionLawNormalized}) and (\ref{eq:DiffEq}) with the 
start condition $\chi_\mathrm{n}=\theta_\mathrm{n}=1$ will give the ray trajectory.

Plasma waves are refracted toward lower plasma densities in the magnetosphere until refraction becomes inefficient due to the decreasing plasma density along its trajectory.
As can be seen in Fig. \ref{Fig_Refraction_Hearts} refraction results in a redistribution of rays; i.e. the rays are no longer
equi-spaced above the emission height and two ``conal components'' of outer rays
and a ``core component'' of inner rays are formed.

\subsection{Calculation of the pulse profiles}

The effect of refraction is quantified in a plot of the final ray direction
(Fig. \ref{Fig_Refr_FRD}).  Here the propagation direction $\theta_f$ at a
height where refraction has become inefficient
is plotted versus $\chi_s$. 
The initial ray position $(r_0, \chi_0)$ corresponds to a value of $\chi_s$ by tracing back the field line from the emission height to $r=1$ (see Fig. \ref{fig:Coordinates}).

Both inside and outside the characteristic
field line cone refraction 
is toward lower plasma densities, which results in a steepening of the final ray direction plot near $\chi_s=\chi_c$.
Inside the plasma cone the rays are refracted toward the magnetic axis and the innermost rays may even intersect the magnetic axis; in that case the final ray direction plot
crosses the line $\theta_f = 0$.

For small values of $\theta_f$, rays originating from 
several discrete values of $\chi_s$ leave the magnetosphere in the same direction and at the 
corresponding pulse longitude different parts of the emission ring can be 
observed simultaneously.
Note that the final ray direction curve for the opposite half of the emission ring is found by mirroring the curve with respect to the line $\theta_f=0$. If the final ray direction plot crosses this line, some parts of both sides of the emission ring have the same $\theta_f$. In that case both sides of the emission ring can be observed simultaneously, if the impact angle $\beta$ is small enough.

If the curve in Fig. \ref{Fig_Refr_FRD} is horizontal at the $\theta_f$ value 
corresponding to the line of sight $(\theta_\mathrm{LOS})$, a large part of the
emission surface is observed simultaneously while if the curve is steep only a 
small part is observed. This means that the observed intensity in the pulse 
profile is proportional to the value of $d\chi_s/d\theta_f$ at 
$\theta_f=\theta_{\mathrm{LOS}}$ which is just an energy conservation argument 
(P2000).

Apart from refraction effects the pulse profile will depend on the intensity 
distribution at the emission height.
If the pair production is somehow
related to the observed coherent microwave radiation (\citealt{RS}), then
similar distributions for the plasma density and the intensity at the
emission height can be expected such as (P2000) 
\be
\label{Eqn_Refr_EmissionRing_Gauss}
W_{r_0} = \exp\left(-\varepsilon\Upsilon\left(\frac{|\chi_s|-\chi_c}{\chi_c}\right)^2\right).
\ee 
This corresponds to an emission ring which peaks at the characteristic field lines
and its thickness is set by $\Upsilon$.
For $\Upsilon=1$ the intensity distribution
follows exactly the plasma density distribution.
The shape is Gaussian as a function of the field line parameter $\chi_s$, but the choice of another parameter (such as the length along the emission surface) is also conceivable. However for simplicity the parameter $\chi_s$ is used. As will be discussed later on, the conclusions do not depend on this choice.

The refraction model is axisymmetric around the magnetic axis, so the
beam-pattern of the pulsar is also axisymmetric around the same axis.  The shape
of observed pulse profiles depends on how the line of sight cuts the
magnetic pole of the star. We only consider the most simple geometry;
i.e. the magnetic axis is orthogonal to the rotation axis
($\alpha=\degrees{90}$) and the line of sight cuts the magnetic pole
centrally (impact angle $\beta=\degrees{90}$).  For this geometry the pulse
longitude $\phi$ is equal to the final ray direction $\theta_f$. Because of
this choice of geometry and the axisymmetry, all the information of the
beam-pattern is in the calculated pulse profiles.  The model itself is
independent of $\alpha$ and $\beta$, only the mapping between $\phi$ and
$\theta_f$ changes.

\begin{figure*}[tb]
\begin{center}
\includegraphics[width=1.99\columnwidth, angle=0]{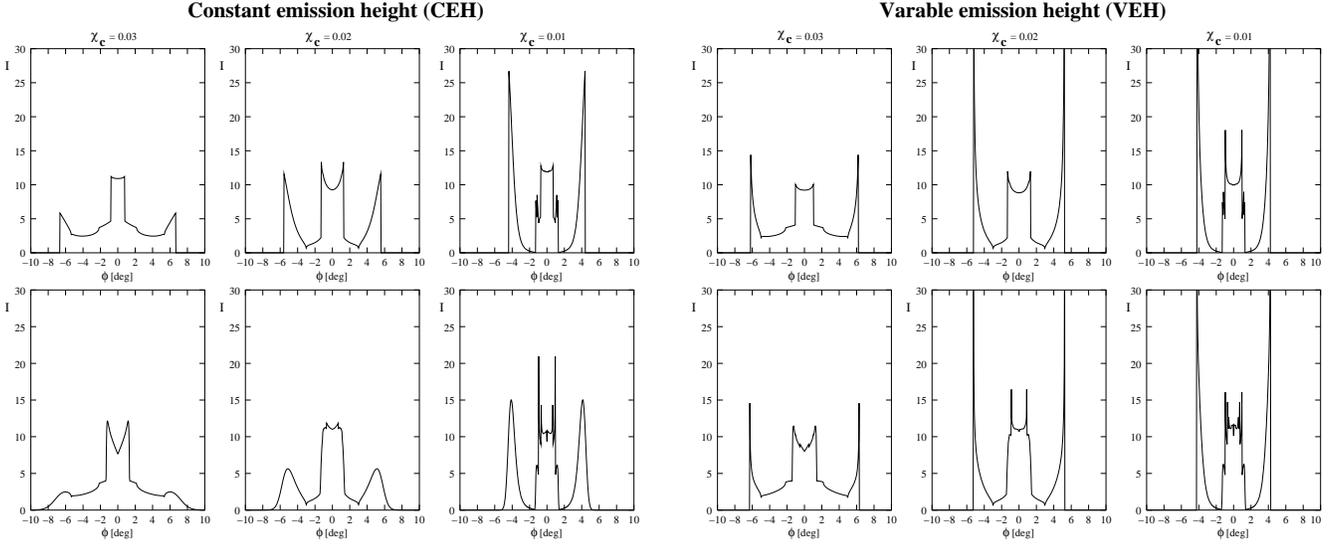}
\caption{\label{Fig_Refr_PulseProfiles}Pulse profiles for different $\chi_c$
(a small $\chi_c$ corresponds to a high observing frequency) for both the
CEH and the VEH model. For the top row $\chi_s$ ranges from $\chi_c/2$ to
$3\chi_c/2$ with $\Upsilon=0$ (all field lines having equal
intensity) and for the bottom row $\chi_s$ ranges from $\chi_c/5$ to
$5\chi_c/2$ with $\Upsilon=1$ (intensity coupled 
to the plasma density).  The other parameters are the same as in
Figs. \ref{Fig_Refraction_Hearts} and \ref{Fig_Refr_FRD}. The scale is such that the integrated intensity is the same for all profiles.} 
\end{center}
\end{figure*}

\subsection{\label{sct:VEH}Variable emission height model}

The model described above may be applied with both a constant (CEH) and a variable
emission height (VEH), but (as we will argue) a VEH is needed to make the model self-consistent. The requirement of a VEH was not met in P2000; it is
the basic conceptual difference between the model presented here and the
P2000 model. Its effect on the pulse profiles turns out to be appreciable,
as discussed below.

The emission height can be derived when a plasma density distribution has
been specified. The plasma density decreases as $r^{-3}$, so the local plasma
frequency decreases away from the star resulting in the excitation of plasma
waves with higher frequencies closer to the star. This results
in rays propagating in a direction which is more aligned with the magnetic axis at the emission
height. But there is another effect involved in the frequency dependence of
the pulse profile morphology: refraction becomes more prominent.

The assumption that both $f_0$ and $\gamma$ are constant implies that plasma
waves of one particular frequency are generated at one particular
equi-plasma density surface (Eqs. \ref{eq:OmegaP} and
\ref{fDefenition}).  Because a transverse plasma density gradient is needed
for refraction, the emission height of a given frequency varies with polar
angle $\chi_0$.  If the magnetic field is dipolar, we have
\be
\label{eq:chis}
\chi_0 = \chi_s\sqrt{r_0}, \ee where $\chi_s$ is shown in Fig.
\ref{fig:Coordinates}.  Combining the plasma density (Eq.
\ref{Eqn_Refr_PlasmaDensity}) with Eqs. (\ref{eq:OmegaP}) and
(\ref{fDefenition}) at the emission height, and using Eq. (\ref{eq:chis})
leads to the following expression for the emission height
\begin{eqnarray}
\nonumber r_0^3 &=& 
R^3\exp\left(-\varepsilon\left(\frac{|\chi_s|-\chi_c}{\chi_c}\right)^2\right)\\
\label{EmissionSurface}R^3 &=& \frac{4\pi\gamma e^2f_0^2N_\star}{m\omega^2}.
\end{eqnarray}

Because $\gamma$ and $f_0$ are assumed to be constant, $R$ should be
constant and $r_0$ is not constant. As noted earlier, the whole problem is
independent of the scale of $r$, so $R$ can be set equal to 1.  The frequency
dependence of the pulse profiles is then in the parameter $\chi_c$ (P2000)
\begin{eqnarray}
\chi_c \propto \omega^{-1/3}.
\end{eqnarray}
The ray trajectory is solved as a function of the distance to the star,
expressed in units of the emission height and different $\chi_c$ correspond
to relative observing frequencies. The physical emission height can be
calculated from Eq. (\ref{EmissionSurface}) when $N_\star$ and $\omega$
are specified.

The emission surface specified by Eq. (\ref{EmissionSurface}) corresponds
to an isodensity surface, so the plasma density distribution has a more
prominent role in this VEH model than in the CEH model.  Apart from
causing refraction, it also determines the shape of the emission surface.

Refraction becomes more severe for the inner and outer rays in the VEH
model, because the plasma gradients are larger at lower emission heights.
Moreover a lower emission height implies that the rays are emitted closer
to, and are initially propagating more aligned with the magnetic axis.  For
the inner rays this means that the rays can intersect the magnetic axis
more easily. For the outer rays there are two counteracting effects. A lower
emission height implies that the rays are refracted in a more outward direction, but at
the same time the rays are also emitted more aligned to the magnetic axis.

\section{Results}

Model calculations of pulse profiles for both a VEH and a CEH are presented
in Fig. \ref{Fig_Refr_PulseProfiles} for the most simple geometry
($\beta=\degrees{0}$ and $\alpha=\degrees{90}$).  For this geometry the
pulse longitude $\phi$ is equal to the final ray direction $\theta_f$.

Observationally the core component behaves differently from conal
components, both in the frequency dependence of its morphology and in its
polarization properties. This is what can be expected from refraction
(P2000), because the core component consists of ``mixed'' rays; i.e. the order of
the beams changes.  This is true for both a VEH and a CEH.

In Fig. \ref{Fig_Refraction_Hearts} one can see that refraction becomes
more prominent for higher frequencies (lower $\chi_c$).  This can also be
seen in Fig. \ref{Fig_Refr_FRD} where the final propagation direction of
rays versus $\chi_s$ is plotted. The curve becomes more complex for lower
$\chi_c$. Besides this refractive effect, a lower emission height implies
smaller propagation angles $\theta$ at the emission height, resulting in
narrower pulse profiles with increasing frequency (decreasing $\chi_c$) in
Fig. \ref{Fig_Refr_PulseProfiles}. This is again true for both a VEH and a CEH.

The pulse profiles in
Fig. \ref{Fig_Refr_PulseProfiles} for a CEH are more spiky than the pulse
profiles presented in P2000. The main reason for this is the higher
resolution of the calculations presented here.

There are three reasons why the profiles, for both a VEH and a CEH, are spiky. First of all,
if the intensity distribution at the emission height is flat, the emission
ring has sharp edges resulting in sharp edges in the pulse profile. This
effect can be reduced by making $\Upsilon$ larger (see Eq. \ref{Eqn_Refr_EmissionRing_Gauss}), as is seen in the bottom row of Fig.
\ref{Fig_Refr_PulseProfiles}. Making the parameter $\Upsilon$ larger results
in the edge of the intensity distribution becoming Gaussian blurred.

The second effect is caused by rays crossing the magnetic axis, so this applies especially to
the core component. This means that at certain pulse longitude both
sides of the emission ring can be seen simultaneously. When the number of
sides visible changes at a particular pulse longitude, a step in intensity
appears in the pulse profile. This effect can again be reduced by increasing
$\Upsilon$ as can be seen in the bottom row of \ref{Fig_Refr_PulseProfiles}.

The last effect contributing to the spikiness of the profiles is a focusing effect. If a large patch of the emission
ring is focused at one pulse longitude, a peak is observed. This focusing
effect corresponds to a horizontal part in the final ray direction curve;
rays emitted at a range of $\chi_s$ are focused to a single $\theta_f$.  At
high frequencies this can be seen for the innermost rays (Fig.
\ref{Fig_Refr_FRD}), causing peaks in the core component.

By comparing the pulse profiles for the VEH and the CEH model in Fig.
\ref{Fig_Refr_PulseProfiles}, the most striking difference is the conal
components. The edges of the VEH profiles are very sharp, and introducing a large $\Upsilon$ will not reduce their sharpness. The reason
can be found in Fig. \ref{Fig_Refr_FRD}. The curves for the VEH model show
a global maximum at $\chi_s\approx1.5$. Because it is a maximum, there is
focusing and because the maximum is global, the peaks occur at the edges
of the profile.

\begin{figure}
\includegraphics[width=0.99\columnwidth, angle=0]{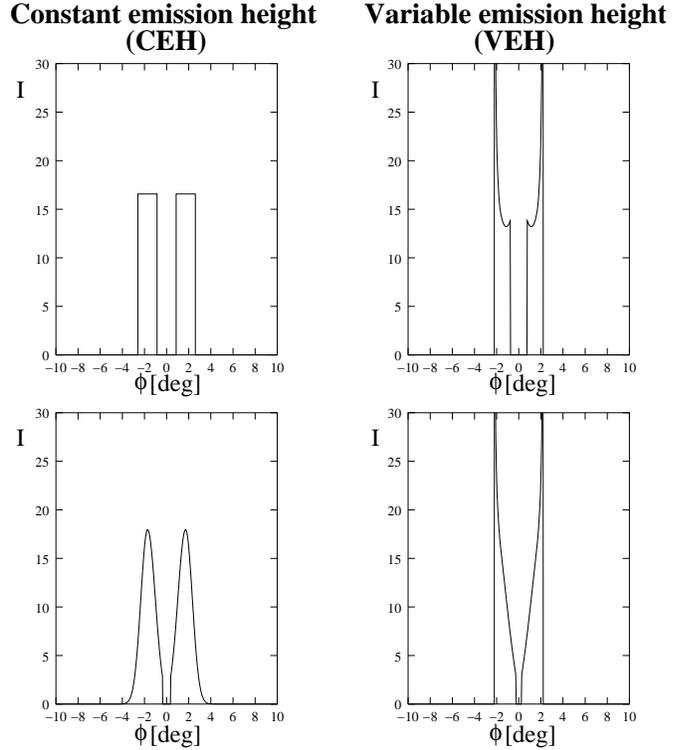}
\caption{Pulse profiles calculated for the case of no refraction.
For the top row $\chi_s$ ranges from $\chi_c/2$ to
$3\chi_c/2$ with $\Upsilon=0$ and for the bottom row $\chi_s$ ranges from $\chi_c/5$ to $5\chi_c/2$ with $\Upsilon=1$. These profiles have been calculated for $\chi_c=0.02$. The other the parameters and the normalization of the profiles are the same as in Fig. \ref{Fig_Refr_PulseProfiles}.}
\label{fig:W}
\end{figure}

As discussed in Sect. \ref{sct:VEH}, there are two counteracting
effects for the outer rays in the VEH model. A lower emission height makes
the outward directed refraction stronger, but the propagation angle
is more aligned with the magnetic axis at the emission height. The reason for
the global maximum is that the latter effect dominates for the outermost
rays.

This focusing effect due to the variable emission height is also visible in the pulse profiles of Fig. \ref{fig:W}, which were calculated without using refraction. The focusing is caused by the geometry of the emission surface, not by the intensity distribution at the emission height (Eq. \ref{Eqn_Refr_EmissionRing_Gauss}). This means that this focusing is independent of the precise form of Eq. (\ref{Eqn_Refr_EmissionRing_Gauss}), and therefore also of the choice to use the field line parameter $\chi_s$ instead of for example the length along the emission surface in this equation.

The edges of the profiles are produced by rays emitted from
$\chi_s\approx1.5$ and the rays are focused, so there should be only very
little radiation produced at $\chi_s\approx1.5$ to avoid the sharp edge.
This means that the emission ring should be very thin, so $\Upsilon$ should
be large.  At high frequencies $\Upsilon$ should be at least $\approx 5$ and
at the lowest frequencies ($\chi_c=0.03$) $\Upsilon$ should be at least
$\approx 15$.  A large $\Upsilon$ physically means that only the middle part
of the emission ring is producing coherent microwave radiation, although the
whole ring is producing streams of particles.  Such a scenario is in
conflict with the expectation that pair production and coherent emission are
related.

A VEH leads to stronger refraction.
Besides introducing a VEH refraction can also be increased by changing the values of $\varepsilon_1$, $\varepsilon_2$, $f_0$ or $\gamma$ in the CEH model. 
Experimenting with a range of values of these parameters did not lead to the formation of the sharp edge of the profiles with a CEH. Therefore the focusing effect is a typical property of the VEH model.

\section{Discussion}

Contrary to the expectation expressed in P2000, the qualitative features of
profile formation turn out to be different for the VEH and the CEH
refraction models. Although the VEH is a physical improvement in the sense
that it makes the emission model self-consistent, the profiles obtained are
less realistic. The model, therefore, needs further improvements before it
can serve as a tool to fit (typical) multifrequency pulsar observations.

The most pronounced difference between the CEH and the VEH model is that for
the VEH model the rays emitted at the outside of the emission surface do not
form the edges of the pulse profile.  The edges of the pulse profiles in the
VEH model are generated by a focusing effect causing the edges to be sharp. If the thickness of the emission ring at the emission height were much thinner than the thickness of the plasma cone at the emission height the sharp edges would disappear, but this seems physically unrealistic.

It must be noted that the results depend strongly on the plasma distribution
adopted. The density profile not only causes refraction, but it also determines the shape of
the emission surface. If the plasma density falls off more slowly than the
Gaussian distribution assumed here, the results may be more realistic
although in that case refraction will be less prominent.

Several other effects could contribute to smoother pulse profiles. There are
probably more frequencies generated at one point in the magnetosphere, so
there would be a $f_0$ range rather than a fixed $f_0$ value.  Also the rays
are not emitted strictly aligned with the magnetic field lines, rather there
will be an elementary beam pattern of finite angular width.  A beam pattern
with a width of $\gamma^{-1}$ can be considerable compared with the pulse
width (for a plasma outflow Lorentz factor $\gamma\approx30$ the beam is
about $\degrees{2}$ wide).  If the outflow Lorentz factor is different for
different field lines, the shape of the emission surface is
changed. Moreover refraction becomes more complex, because it depends on
gradients of the $\gamma$ factor as well as gradients in the plasma density
(\citealt{Barnard86}).

\appendix
\section{Analytical solution of the normalized dispersion relation}

The normalized dispersion relation (\ref{DispersionLawNormalized}) is of the third degree in $\eta$, so the analytical solution of $\eta$ is given by the 
cubic. The solution for the superluminous O-mode (the solution with
positive $\eta$ at the emission height) is
\be
\label{Eqn_Reft_EtaAnalytisch}
\eta=s_++s_--\frac{a_2}{3},
\ee
where
\begin{eqnarray}
\begin{array}{lllll}
s_\pm&=&\left(r\pm\sqrt{q^3+r^2}\right)^{1/3}\\
q&=&\frac{a_1}{3}-\frac{a_2^2}{9}\\
r&=&\frac{1}{6}(a_1a_2-3a_0)-\frac{a_2^3}{27}
\end{array}
\end{eqnarray}
and
\begin{eqnarray}
\nonumber a_2 &=& 2+a_0\\
a_1 &=& 1-4\frac{N}{f_0^2}+2a_0\\
\nonumber a_0 &=& -\frac{9}{4}\chi_0^2\gamma^2(\theta_\mathrm{n}-\chi_\mathrm{n})^2\mathrm{.}
\end{eqnarray}

\acknowledgements We thank Svetlana Petrova for making available her code and for her valuable comments as referee of this article. We also thank her as well as Joeri van Leeuwen, Ramachandran and John Barnard for constructive discussions.


\end{document}